\input epsf

\newfam\msbfam
\font\twlmsb=msbm10 at 12pt
\font\eightmsb=msbm10 at 8pt
\font\sixmsb=msbm10 at 6pt
\textfont\msbfam=\twlmsb
\scriptfont\msbfam=\eightmsb
\scriptscriptfont\msbfam=\sixmsb
\def\cj{\fam\msbfam}

\def\C{{\cj C}}

\def\R{{\cj R}}

\centerline{\bf AHARONOV-BOHM EFFECT VS. DIRAC MONOPOLE: A-B $\Leftrightarrow$ D} 

\

\centerline{Miguel Socolovsky{$^*$}}

\

\centerline{\it  Instituto de Ciencias Nucleares, Universidad Nacional Aut\'onoma de M\'exico}
\centerline{\it Circuito Exterior, Ciudad Universitaria, 04510, M\'exico D. F., M\'exico} 

\

{\bf Abstract.} {\it In the context of fiber bundle theory, we show that the existence of the Aharonov-Bohm connection implies the existence and uniqueness of the Dirac connection.}

\

{\bf Keywords}: {\it Aharonov-Bohm effect, Dirac monopole, fiber bundles}

\

PACS numbers: 02.40.-k, 02.40.Re, 03.65. Vf, 03.65.-w

\

{\bf 1. Introduction} 

\

As is well known [1], the Aharonov-Bohm effect ($A-B$) [2,3] can be described in the trivial $U(1)$-bundle $$\xi_{A-B}:U(1)\hookrightarrow\C^*\times U(1)\buildrel {pr_1}\over\longrightarrow \C^*, \ \ pr_1(z,\mu)=z \eqno{(1)}$$ ($\C^*=\C\setminus\{0\}$, $\mu=e^{i\varphi}$, $\varphi\in [0,2\pi)$), while the hypotetical $g={{1}\over{2}}$ magnetic charge or Dirac monopole ($D$) [4,5] can be described in the non trivial $U(1)$-bundle [6] $$\xi_D: U(1)\hookrightarrow S^3\buildrel {\pi_H}\over\longrightarrow S^2\eqno{(2)}$$ (Hopf bundle, [7]), where $\pi_H$ is the Hopf map $$\pi_H(z_1,z_2)=\{\matrix{z_1/z_2, \ \ z_2\neq 0\cr \infty, \ \ z_2=0\cr}\eqno{(3)}$$ with $$S^3=\{(z_1,z_2)\vert \ \ \vert z_1\vert^2+\vert z_2\vert^2=1\}\subset\C^2\eqno{(4)}$$ and $$\Phi:S^2\subset\R^3\buildrel {\cong}\over\longrightarrow \C\cup\{\infty\},\ \ \Phi(x_1,x_2,x_3)=\{\matrix{{{x_1+ix_2}\over{1+x_3}}, \ \ (x_1,x_2,x_3)\neq (0,0,-1)\cr \infty, \ \ (x_1,x_2,x_3)=(0,0,-1)\cr}.\eqno{(5)}$$ In terms of the Euler angles in $\R^3$, $\chi,\varphi\in [0,2\pi)$, $\theta\in [0,\pi]$, the $u(1)=Lie(U(1))$-valued $D$ connection on $S^3$ is given by [8] $$\omega_D={{i}\over{2}}(d\chi+cos \theta \ d\varphi)\eqno{(6)}$$ with $D$ potentials on $S^2$ $${A_D}_{\pm}=\mp{{i}\over{2}}(1-cos \theta) \ d\varphi,\eqno{(7)}$$ and curvature $F_D=d\omega_D={{i}\over{2}}sin \theta \ d\theta\wedge d\varphi$: $(-i)\times$ the magnetic field of the monopole, while the $A-B$ potentials on $\C^*$ (and global connection $A_{A-B}$ on $\C^*\times U(1)$ since $\xi_{A-B}$ is trivial) are given by [1] $${A_{A-B}}_\pm=\mp{{i}\over{2}}d\varphi=\mp{{i}\over{2}}{{X_1dX_2-X_2dX_1}\over{X_1^2+X_2^2}}\eqno{(8)}$$ with $z=X_1+iX_2\in\C^*$ and $X_1$, $X_2$ the Cartesian coordinates on ${\R^2}^*\cong\C^*$; clearly, ${A_{A-B}}_\pm$ are closed ($A_{A-B}$ is flat in its domain of definition, $z\neq 0$) but not exact 1-forms. 

\

From (7) and (8), $${A_D}_{\pm}\vert_{\theta=\pi/2}={A_{A-B}}_\pm\eqno{(9)}$$ which, in the context of bundle theory, tells us that the existence of the Dirac monopole implies the existence of the $A-B$ effect ($``D\Rightarrow A-B"$). The same conclusion has been arrived at in ref. [9], where the close relation between both phenomena was exhibited by showing that the $A-B$ bundle is equivalent (isomorphic) to the pull-back of the $D$ bundle by the inclusion $\iota:\C^*\to\C\cup\{\infty\}$, $\iota(z)=z$, between the corresponding base spaces: $$\xi_{A-B}\cong\iota^*(\xi_D).\eqno{(10)}$$ 

\

This fact immediately raises the question for the inverse implication, namely, if the existence of the $A-B$ effect implies, at least in the present mathematical sense, the existence of the Dirac monopole [10]. These monopoles, though yet not found in Nature, are predicted by grand unified [11] and string [12] theories. The purpose of the present note is to answer affirmatively the above question. 

\

{\bf 2. Pull-back of the $A-B$ potentials}

\

If $N=(0,0,1)$ and $S=(0,0,-1)$ are the north and south poles of $S^2$, then $$\pi_H^{-1}(\{N,S\})=\{(z_1,0), \ \vert z_1\vert=1\}\cup\{(0,z_2), \ \vert z_2\vert=1\}\cong S^1\times S^1=T^2,\eqno{(11)}$$ the 2-torus. If we ``truncate" the $\xi_D$ bundle by defining the $U(1)$-bundle $$\hat{\xi_D}:U(1)\hookrightarrow S^3\setminus T^2\buildrel{\pi_H\vert}\over\longrightarrow S^2\setminus\{N,S\}\cong\C^*,\eqno{(12)}$$ the inclusion $\iota:\C^*\to S^2$ becomes the identity $Id_{\C^*}$ and we have the bundle map given by Diagram 1: $$\matrix{(\C^*\times U(1))\times U(1) & \buildrel{\bar{\iota}\times Id_{U(1)}}\over\longrightarrow & (S^3\setminus T^2)\times U(1)\cr \psi_0\downarrow & & \downarrow\psi_D\vert\cr \C^*\times U(1) & \buildrel{\bar{\iota}}\over\longrightarrow & S^3\setminus T^2\cr  pr_1\downarrow & & \downarrow\pi_H\vert  \cr \C^* & \buildrel{Id_{\C^*}}\over\longrightarrow & \C^*\cr}$$ \centerline{\it Diagram 1} 

\

where $\psi_0$ and $\psi_D\vert$ are the right actions of $U(1)$ on the corresponding total spaces, $$\bar{\iota}(z,\mu)={{(z,1)\mu}\over{\vert\vert(z,1)\vert\vert}},\eqno{(13)}$$ and $\vert$ denotes the corresponding restrictions. It is clear that the ``transitions" from $\omega_D$ and ${A_D}_\pm$ to the restrictions $\omega_D\vert$ and ${A_D}_\pm\vert$ respectively on $S^3\setminus T^2$ and $\C^*$ are continuous, since they amount to the restriction of the domain of $\theta$ from $[0,\pi]$ to $(0,\pi)$.

\

Defining Hopf coordinates [13] $\{\eta,\xi_1,\xi_2\}$ on $S^3$: $$(z_1,z_2)=(e^{i\xi_1}sin \ \eta,e^{i\xi_2}cos \ \eta), \ \eta\in[0,\pi/2], \ \xi_1,\xi_2\in[0,2\pi]\eqno{(14)}$$ we obtain $$\pi_H\vert(\eta,\xi_1,\xi_2)=e^{i(\xi_1-\xi_2)}tg \ \eta, \eqno{(15)}$$ with $\eta\in(0,\pi/2)$, which allows us to construct the pull-back $\beta\in \Omega^1(S^3\setminus T^2;u(1))$ of ${A_{A-B}}_\pm\in \Omega^1(\C^*;u(1))$ by $\pi_H\vert^*$: $$\pmatrix{\beta_\eta\cr \beta_{\xi_1}\cr \beta_{\xi_2}\cr}=\pm\pmatrix{{{\partial}\over{\partial\eta}}(X_1\circ\pi_H\vert) & {{\partial}\over{\partial\eta}}(X_2\circ\pi_H\vert)\cr {{\partial}\over{\partial\xi_1}}(X_1\circ\pi_H\vert) & {{\partial}\over{\partial\xi_1}}(X_2\circ\pi_H\vert)\cr {{\partial}\over{\partial\xi_2}}(X_1\circ\pi_H\vert) & {{\partial}\over{\partial\xi_2}}(X_2\circ\pi_H\vert\cr}\pmatrix{{A_{A-B}}_{1\pm}\cr {A_{A-B}}_{2\pm}\cr}$$ $$=\pm\pmatrix{{{cos(\xi_1-\xi_2)}\over{cos^2\eta}} & {{sin(\xi_1-\xi_2)}\over{cos^2\eta}}\cr -sin(\xi_1-\xi_2)tg \ \eta & cos(\xi_1-\xi_2)tg \ \eta\cr sin(\xi_1-\xi_2)tg \ \eta & -cos(\xi_1-\xi_2)tg \ \eta}\pmatrix{{A_{A-B}}_{1\pm}\cr {A_{A-B}}_{2\pm}\cr}=\pmatrix{0\cr i/2\cr -i/2\cr}\eqno{(16)}$$ i.e. $$\beta={{i}\over{2}}(d\xi_1-d\xi_2).\eqno{(17)}$$ From the relation between Hopf coordinates and Euler angles, $$(e^{i\xi_1}sin \ \eta,e^{i\xi_2}cos \ \eta)=(e^{{{i}\over{2}}(\varphi+\chi)}cos(\theta/2),e^{{{i}\over{2}}(\varphi-\chi)}sin(\theta/2))\eqno{(18)}$$ one obtains $$\beta={{i}\over{2}}d\chi\eqno{(19)}$$ i.e. $$\pi_H\vert^*({A_{A-B}}_\pm)=\omega_D\vert(\theta=\pi/2).\eqno{(20)}$$

\

{\bf 3. Push-forward of the $A-B$ connection}

\

The same relation between $A_{A-B}$ and $\omega_D\vert(\theta=\pi/2)$ can be arrived at through the more direct path of pushing forward horizontal spaces of $A_{A-B}$ in $\xi_{A-B}$ into horizontal spaces of $\omega_D\vert(\theta=\pi/2)$ in $\hat{\xi_D}$. Since $Id_{\C^*}$ is a diffeomorphism and $Id_{U(1)}$ is a group homomorphism (isomorphism), we are in the conditions of Proposition 6.1. in ref. [14]: given $A_{A-B}$ in $\xi_{A-B}$ there {\it exist} and is {\it unique} a connection $\omega$ in $\hat{\xi_D}$ such that the horizontal subspaces of $A_{A-B}$ in $\C^*\times U(1)$ are mapped into the horizontal subspaces of $\omega$ in $\hat{\xi_D}$ by $d\bar{\iota}\equiv \bar{\iota}_*$. Here, we shall explicitly prove this fact and find that $\omega=\omega_D\vert(\theta=\pi/2)$. 

\

At any point $(z,e^{i\varphi})$ of $\C^*\times U(1)$, the horizontal space of $A_{A-B}$ is the kernel of (8). So, $(X_1dX_2-X_2dX_1)(V_1{{\partial}\over{\partial X_1}}+V_2{{\partial}\over{\partial X_2}})=X_1V_2-X_2V_1=0$ implies $$V_2={{X_2}\over{X_1}}V_1 \ for \ \ X_1\neq 0, \ and \  V_1=0 \ for \ X_1=0.\eqno{(21)}$$ Since $$T_{(z,e^{i\varphi})}(\C^*\times U(1))=T_z\C^*\oplus T_{e^{i\varphi}}U(1)=\C\oplus\{te^{i(\varphi+\pi/2)}\}_{t\in\R}, \eqno{(22)}$$ the horizontal vectors at $(z,e^{i\varphi})$ are given by $$V\equiv(V_1,V_2,V_\varphi)=\{\matrix{(V_1,{{X_2}\over{X_1}}V_1,ite^{i\varphi}), \ X_1\neq 0\cr (0,V_2,ite^{i\varphi}), \ X_1=0\cr}.\eqno{(23)}$$ On the other hand, from the definition of $\bar{\iota}$ in eq. (13) and the definition of the Hopf coordinates on $S^3$, eq. (14), one obtains $$\bar{\iota}(z,e^{i\varphi})=\bar{\iota}(X_1+iX_2,e^{i\varphi})\cong \bar{\iota}(X_1,X_2,e^{i\varphi})=(\eta(X_1,X_2,\varphi),\xi_1(X_1,X_2,\varphi),\xi_2(X_1,X_2,\varphi))$$ $$=(tg^{-1}(\sqrt{X_1^2+X_2^2}),tg^{-1}({{X_2cos \ \varphi+X_1sin \ \varphi}\over{X_1cos \ \varphi-X_2sin \ \varphi}}),\varphi),\eqno{(24)}$$ leading to $W=\bar{\iota}_*(V)$ with components $$\pmatrix{W_\eta\cr W_{\xi_1}\cr W_{\xi_2}\cr}=\pmatrix{{{\partial\eta}\over{\partial X_1}} & {{\partial\eta}\over{\partial X_2}} & {{\partial\eta}\over{\partial\varphi}}\cr {{\partial\xi_1}\over{\partial X_1}} & {{\partial\xi_1}\over{\partial X_2}} & {{\partial\xi_1}\over{\partial\varphi}}\cr {{\partial\xi_2}\over{\partial X_1}} & {{\partial\xi_2}\over{\partial X_2}} & {{\partial\xi_2}\over{\partial\varphi}}\cr}\pmatrix{V_1\cr V_2\cr V_\varphi\cr}=\pmatrix{{{X_1}\over{(tg \ \eta)(1+tg^2\eta)}} & {{X_2}\over{(tg \ \eta)(1+tg^2\eta)}} & 0\cr -{{X_2}\over{tg^2\eta}} & {{X_1}\over{tg^2\eta}} & 1\cr 0 & 0 & 1\cr}\pmatrix{V_1\cr V_2\cr V_\varphi\cr}.\eqno{(25)}$$ The relation between Hopf and Euler coordinates: $$tg \ \eta=cotg({{\theta}\over{2}}), \ \xi_1={{\varphi+\chi}\over{2}}, \ \xi_2={{\varphi-\chi}\over{2}}, \eqno{(26)}$$ allows to write $$\omega_D(\eta,\xi_1,\xi_2)={{i}\over{1+tg^2\eta}}(tg^2\eta \ d\xi_1-d\xi_2).\eqno{(27)}$$ In particular, $$\omega_D(\pi/4,\xi_1,\xi_2)=\omega_D(\theta=\pi/2).\eqno{(27a)}$$ The horizontal space at any point $(\eta,\xi_1,\xi_2)\in S^3\setminus T^2$, being the kernel of $\omega_D\vert$, turns out to be $$H_{(\eta,\xi_1,\xi_2)}=\{(v_\eta,v_1,(tg^2\eta)v_1), \ v_\eta,v_1\in\R, \ \eta\in(0,\pi/2)\}.\eqno{(28)}$$ In particular, $$H_{(\pi/4,\xi_1,\xi_2)}=\{(v_{\pi/4},v_1,v_1), \ v_{\pi/4},v_1\in\R\}.\eqno{(28a)}$$ For $X_1\neq 0$, (25) leads to $$\pmatrix{W_\eta\cr W_{\xi_1}\cr W_{\xi_2}\cr}=\pmatrix{{{V_1tg \ \eta}\over{X_1(1+tg^2\eta)}}\cr ite^{i\varphi}\cr ite^{i\varphi}}\eqno{(29)}$$ while for $X_1=0$, (25) leads to $$\pmatrix{W_\eta\cr W_{\xi_1}\cr W_{\xi_2}\cr}=\pmatrix{{{X_2V_2}\over{(tg \ \eta)(1+tg^2\eta)}}\cr ite^{i\varphi}\cr ite^{i\varphi}},\eqno{(30)}$$ which belong to $H_{(\pi/4,\xi_1,\xi_2)}$. So, horizontal spaces of $A_{A-B}$ are mapped into horizontal spaces of $\omega_D\vert(\theta=\pi/2)$. 

\

{\bf 4. Unique determination of $\omega_D$}

\

By symmetry reasons, the {\it unique} $\theta$-dependent extensions $\hat{\omega}$ of $\omega$ are of the form $sin \ \theta \ d\theta$, $sin \ \theta \ d\varphi$, $cos \ \theta \ d\varphi$, and $cos \ \theta \ d\theta$. The first two lead to $\hat{\omega}(\theta=\pi/2)={{i}\over{2}}(d\chi+d\theta)$ or ${{i}\over{2}}(d\chi+d\varphi)$ which are different from $\omega_D\vert(\theta=\pi/2)$, while the fourth one leads to $\hat{\omega}={{i}\over{2}}(d\chi+dsin \ \theta)={{i}\over{2}}d\chi^\prime$, with $\chi^\prime=\chi+sin \ \theta$, which is the same as $\omega$. So, the unique $\theta$-dependent extension of $\omega$ is the restriction to $S^3\setminus T^2$ of the Dirac connection: $$\hat{\omega}(\theta)=\omega_D\vert(\theta).\eqno{(31)}$$ Since $\omega$ in $\hat{\xi_D}$ is uniquely determined by $A_{A-B}$ in $\xi_{A-B}$, its $\theta$-dependent extension $\hat{\omega}$ is unique, and, as previously mentioned, the transition from $\theta\in(0,\pi)$ to $\theta\in[0,\pi]$ is continuous, then $A_{A-B}$ uniquely determines $\omega_D$. This ends the proof of the existence and uniqueness of the $D$ connection from the existence of the $A_{A-B}$ connection.

\

{\bf 5. Final comment}

\

The present note does not claim to prove the physical existence of the Dirac monopole, but only to reinforce this idea by showing that, at the mathematical level, in particular in the context of fiber bundle theory, the Aharonov-Bohm connection, relevant to the physically observed $A-B$ effect, implies the existence and uniqueness of the connection which represents the till now hypotetical Dirac charge.

\

{\bf Acknowledgments}

\

The author thanks IAFE (UBA-CONICET), Argentina, for hospitality.

\

{\bf References}

\

[1] M.A. Aguilar, M. Socolovsky, {\it Aharonov-Bohm Effect, Flat Connections, and Green's Theorem}, Int. Jour. Theor. Phys. {\bf 41}, (2002), 839-860.

\

[2] Y. Aharonov, D. Bohm, {\it Significance of electromagnetic potentials in the quantum theory}, Phys. Rev. {\bf 115}, (1959), 485-491.

\

[3] R. G. Chambers, {\it Shift of an electron interference pattern by enclosed magnetic flux}, Phys. Rev. Lett. {\bf 5}, (1960), 3-5.

\

[4] P.A.M. Dirac, {\it Quantized Singularities in the Electromagnetic Field}, Proc. Roy. Soc. {\bf A133}, (1931), 60-72.

\

[5] P.A.M. Dirac, {\it The Theory of Magnetic Poles}, Phys. Rev. {\bf 74}, (1948), 817-830.

\

[6] M. Socolovsky, {\it Spin, Monopole, Instanton and Hopf Bundles}, Aportaciones Matem\'aticas, Notas de Investigaci\'on {\bf 6}, Soc. Mat. Mex., (1992), 141-164. 

\

[7] H. Hopf, {\it Uber die Abbildungen der dreidimensionalen Sphare auf die Kugelflache}, Math. Ann. {\bf 104}, (1931), 637-665.

\

[8] A. Trautman, {\it Solutions of the Maxwell and Yang-Mills Equations Associated with Hopf Fibrings}, Int. Jour. Theor. Phys. {\bf 16}, (1977), 561-565.

\

[9] M. Socolovsky, {\it Aharonov-Bohm effect, Dirac monopole, and bundle theory}, Theor. Phys. {\bf 3}, Nr. 3 (sept. 2018, in press).

\

[10] M. Socolovsky, {\it Sequence of maps betwwen Hopf and Aharonov-Bohm bundles}, Theor. Phys. {\bf 3}, Nr. 4 (dec. 2018, in press).

\

[11] J. Preskill, {\it Magnetic Monopoles}, Ann. Rev. Nucl. Part. Sci. {\bf 34}, (1984), 461-530.

\

[12] J. Polchinski, {\it Monopoles, Duality, and String Theory}, Int. Jour. Mod. Phys., {\bf A19}, (2004), 145-156.

\

[13] G.L. Naber, Topology, Geometry, and Gauge Fields. Foundations, Springer-Verlag, New York, (1997), pp. 11-20.

\

[14] S. Kobayashi, K. Nomizu, Foundations of Differential Geometry, Vol. I, John Wiley, New York, (1963), pp. 79-81.

\

\

\

\

$^*$ e-mail: socolovs@nucleares.unam.mx

\end